Enclosure: gmuon  [30,942 bytes]
----------------------------------------------------------------------
\documentstyle[12pt]{article}
\def\lapx{\,\,\lower 2pt \hbox{$\buildrel<\over{\scriptstyle{\sim}}$}\,\,}
\def\gapx{\,\,\lower 2pt \hbox{$\buildrel>\over{\scriptstyle{\sim}}$}\,\,}
\begin{document}
\setcounter{page}{0}
\thispagestyle{empty}
\begin{flushright}
UTPT-93-7 \\
hep-ph/9304264 \\
\end{flushright}
\phantom{1}
\vspace{10mm}
\begin{center}
\Large
Hadronic contribution to the photon vacuum polarization: \\
a theoretical estimate \\
\normalsize
\vspace{15mm}
B. Holdom$^1$, Randy Lewis$^1$ and Roberto R. Mendel$^2$ \\
\vspace{5mm}
{\it $^1$Department of Physics, University of Toronto,} \\
{\it Toronto, Ontario, CANADA~~M5S~1A7} \\
{\it $^2$Department of Applied Mathematics, University of Western Ontario,} \\
{\it London, Ontario, CANADA~~N6A~5B7} \\
\vspace{20mm}
ABSTRACT\\
\end{center}
A simple model for the hadronic contribution to the photon vacuum polarization
function $\Pi_{had}(q^2)$, for spacelike momenta, is presented.  For small
momenta, the two loop contribution from the pseudoscalar meson octet is
computed from the chiral Lagrangian.  The light quark contribution (which at
low momentum gives the ${\cal O}(q^6)$ counterterm in the chiral Lagrangian)
is calculated within a relativistic constituent quark model incorporating the
momentum dependence of the quark mass.  The perturbative gluons of QCD are
included in a standard fashion.
The total result is close to an estimate of $\Pi_{had}(q^2)$ that is obtained
directly from $e^+e^-\rightarrow hadrons$ data.  We further use
our results for $\Pi_{had}(q^2)$ to calculate the ${\cal O}(e^4)$ hadronic
contribution to lepton magnetic moments and to calculate $\alpha_{QED}(M_Z^2)$.
A simpler model of constituent quarks with momentum independent masses gives
less favourable results. \\
\vspace{10mm}
\begin{flushleft}
To appear in Z. Phys. C.
\end{flushleft}
\newpage

\section*{Introduction}

The photon vacuum polarization function, $\Pi(q^2)$,where
\begin{equation}
i\Pi_{\mu\nu}(q) = ie^2\Pi(q^2)[q^2g_{\mu\nu}-q_{\mu}q_{\nu}]
\end{equation}
plays an important role in any high precision QED observable, e.g. the lepton
anomalous magnetic moments $a_l$ ($l = e,{\mu},{\tau}$), atomic energy levels,
$e^+e^- \rightarrow e^+e^-$ scattering.  For many practical applications, one
requires $\Pi(q^2)$ for spacelike momenta only ($q^2<0$) so we will specialize
to this region, where $\Pi(q^2)$ is a smooth and real function.  We define our
subtraction point such that $\Pi(0)=0$.

The largest source of uncertainty lies in the hadronic contribution,
$\Pi_{had}(q^2)$. The most reliable estimates of $\Pi_{had}(q^2)$ that are
available\cite{Burkhardt} use experimental data for
\begin{equation}\label{Rratio}
R(s) = \frac{{\sigma}(e^+e^- \rightarrow \gamma^* \rightarrow hadrons)}
            {{\sigma}(e^+e^- \rightarrow \gamma^* \rightarrow \mu^+\mu^-)}
\end{equation}
in conjunction with dispersion relations.  A simple analytic fit to this
estimate is given in \cite{Burkhardt}, and we will compare our result
to this fit.

{}From the theoretical viewpoint, $\Pi_{had}(q^2)$ is in principle calculable
in terms of the (current) quark mass parameters and a QCD scale (e.g.
$\Lambda_{\overline{MS}}$).  Our approach is a less ambitious one, namely to
extract $\Pi_{had}(q^2)$ (and other related physical observables) from a
simple QCD-inspired model.  We will show that a simple picture can go a long
way towards providing some theoretical understanding of $\Pi_{had}(q^2)$.

The behavior of $\Pi_{had}(q^2)$ at both low and high momenta is constrained by
QCD.  The high momentum description is provided by perturbative QCD.
The low momentum dependence receives an important contribution
from pseudoscalar meson loops, and this can be calculated using standard
techniques of chiral Lagrangians.  The relevant quantity occurs at
${\cal O}(q^6)$ in the low energy derivative expansion, and thus
an associated two-loop calculation must be performed.  Knowledge of an
${\cal O}(q^6)$ counterterm is also necessary, and at present any estimate of
this quantity is model dependent.

In this paper we shall be using a relativistic constituent quark model of the
three light quarks to smoothly interpolate between the low and high energy QCD
contributions to $\Pi_{had}(q^2)$.  This is a gauged nonlocal constituent (GNC)
quark model which incorporates the momentum dependence of the quark mass as a
natural regulator.  The
pion decay constant and all other quantities appearing in the chiral
Lagrangian to ${\cal O}(q^4)$ have been expressed in terms of this mass
function.  In particular the standard quantities\cite{G&L} $L_1$, $L_2$, $L_3$,
$L_9$, and $L_{10}$ are well described in terms of one parameter (denoted by
$A$ below) appearing in the mass function\cite{GNC}.  Reasonable values for the
other $L_i$'s and the current quark masses are obtained as well\cite{HTV}.  The
model has also been successfully applied to certain other quantities, the pion
electromagnetic form factor and the vector-minus-axial two point function,
beyond ${\cal O}(q^4)$\cite{VMD}.  There has also been a recent
discussion\cite{Ball} of the advantages of our nonlocal regularization in the
context of anomalous processes.  These results suggest that some of the more
essential aspects of nonperturbative QCD
are accounted for by the momentum dependence of the quark mass.

There are two GNC quark loop diagrams which contribute to $\Pi_{had}(q^2)$.
At high momenta the effects of the constituent quark mass become negligible
and the GNC diagrams smoothly approach the naive one-quark-loop contribution
to $\Pi_{had}(q^2)$. We add perturbative QCD corrections to the model
description of the three light quarks.  More precisely we add the hard
gluonic corrections beginning at ${\cal O}(\alpha_s)$ as extracted from a
standard perturbative QCD calculation of $\Pi_{had}(q^2)$.
The contributions to $\Pi_{had}(q^2)$ from the $c$ and $b$ quarks
will be represented completely by perturbative QCD.

We will consider the implications that our calculated $\Pi_{had}(q^2)$ has for
the hadronic contributions to the muon and tau magnetic moments.  Of the two,
the muon magnetic moment is less affected by the various uncertainties.  We
will also consider the QED running coupling at large momentum,
$\alpha_{QED}(M_Z^2)$.

Before discussing the various components of our calculation in more detail, we
present the main results in figs.~\ref{plot1} and \ref{plot2}.  The effects of
the meson loops and the hard gluonic corrections are shown separately.  They
are added to the GNC quark contribution to produce our total result for
$\Pi_{had}(q^2)$.  We find that the total result is quite close to the
experimentally-based estimate\cite{Burkhardt}.  We note as well that for
small $-q^2$, the GNC quark-loop contribution is significantly larger than a
more naive constituent quark model with momentum independent masses, for
example with the typical values $m_u=m_d=330$ MeV and $m_s=550$ MeV.  This
comparison is shown in fig.~\ref{plot3}.

Although we divide our calculation into various parts, we stress that we
are consistently using one model to describe $\Pi_{had}(q^2)$ for all
spacelike momenta.  This is the GNC model with perturbative gluonic effects
added.  The GNC model includes the pseudoscalar mesons,
and as we have said, it nicely reproduces the standard chiral Lagrangian of
low energy QCD.  We will therefore describe our calculation of
$\Pi_{had}(q^2)$ at low $-q^2$ using the language of chiral Lagrangians.  In
the model all meson dynamics, including the meson kinetic terms, are
generated through quark loops.  At higher energies the compositeness of
these mesons will become evident via form factors, and their further
contribution to $\Pi_{had}(q^2)$ will be damped out.  We will enforce this
feature of the model by cutting off the meson-loop contribution at a scale
of order $m_\rho$.

\section*{Pseudoscalar mesons}

For $-q^2$ much less than $m_\rho$,
$\Pi_{had}(q^2)$ can be evaluated from the $SU(3)_L \times SU(3)_R$ chiral
Lagrangian with explicit symmetry breaking terms.  The one-meson-loop
contribution corresponds to a calculation at ${\cal O}(q^4)$ in the low energy
expansion and the counterterm at this order is eliminated by the definition
${\Pi}_{had}(0)=0$.  The leading $q^2$ behavior of $\Pi_{had}(q^2)$ requires a
calculation at ${\cal O}(q^6)$, and this corresponds to a two-loop
calculation.  It is this piece which contains the leading logarithms
$ln({\mu}/m_{\pi})$ and $ln({\mu}/m_K)$.  We
will add this two-loop piece to the finite one-loop contribution, and the
result should
provide a good approximation for $\Pi_{had}(q^2)$ at low $-q^2$.
All diagrams with more than two meson loops are of ${\cal O}(q^8)$ or higher,
and are therefore neglected.

We will follow the definition of dimensional regularization counterterms
($L_i^r({\mu})$) given in \cite{G&L}.  We do not enforce the
${\cal O}(q^2)$ equations of motion, so we must reinstate two terms\cite{HTV}
that
are usually removed by the equations of motion.  In the end our result can be
expressed in terms of the $L_i^r({\mu})$'s of \cite{G&L} as well as one
new parameter, $C^r({\mu})$, which is the ${\cal O}(q^6)$ counterterm.
\begin{eqnarray}\label{2loop}
\Pi_{had}(q^2) & = & {\Pi_{had}^{(4)}(q^2) + \Pi_{had}^{(6)}(q^2)} \\
\Pi_{had}^{(4)}(q^2) & = & \frac{1}{3(4{\pi})^2}\left[
              \frac{2}{3}
                   -\left(1-\frac{4m_\pi^2}{q^2}\right)\int_0^1dz~
                   ln\left(1-z(1-z)\frac{q^2}{m_\pi^2}\right) +
                   \right. \nonumber \\
  & &          \hspace{14.8mm} \left.\frac{2}{3}
                   -\left(1-\frac{4m_K^2}{q^2}\right)\int_0^1dz~
                   ln\left(1-z(1-z)\frac{q^2}{m_K^2}\right) \right]
                   \label{2loop4} \\
\Pi_{had}^{(6)}(q^2) & = &
                 {\frac{q^2}{3(4{\pi}f_0)^2}\left[C^r({\mu})
                 +T_1({\mu})+T_2+T_3(q^2)+T_4(q^2)\right]}\label{2loop6}
\end{eqnarray}
$f_0$ is the pseudoscalar decay constant in the chiral limit, and explicit
expressions for the $T_i$'s are given in the appendix.  We require
that the $\mu$ dependence of $C^r({\mu})$ cancels that of $T_1({\mu})$.

To obtain the numerical value of $\Pi_{had}(q^2)$ for small $-q^2$, we need
only the values of the coefficients $C^r({\mu})$ and $L_i^r({\mu})$.  We will
choose $\mu=m_\rho$ and take the experimental values of the $L_i^r(m_{\rho})$'s
from \cite{Ecker}.  We will give below the value of $C^r(m_{\rho})$ from the
GNC quark model.

But to isolate a purely meson-loop contribution to $\Pi_{had}(q^2)$
we will remove the $C^r(m_{\rho})$ from (\ref{2loop6}).  This
term will effectively be included in the low momentum behavior
of the quark-loop graphs of the GNC quark model.  The remaining terms in
(\ref{2loop6}) represent the two-loop meson contribution.  We will allow this
plus the one-loop result to contribute to the growth of $\Pi_{had}(q^2)$ up to
$-q^2=m_{\rho}^{2}$.  On these scales there may be substantial error in a two
loop calculation; we consider the implications of this below.

The one and two loop meson contributions are displayed in fig.~\ref{plot1} and
compared to the quark-loop contribution.  The latter dominates even though the
contributions at low $-q^2$ are inversely proportional to the mass squares of
the quark and pion.  (The relative enhancement of the quark loops is due
mainly to the colour and spin degrees of freedom.)
But the meson-loop contribution is still significant.
We find an uncertainty of 20\% in the meson-loop contribution at $m_\mu^2$
due to the uncertainty in the $L_i^r$'s.

The fact that the one-loop meson contribution is smaller than the two-loop
contribution is consistent with the dominance of the $\rho$ meson.  The $\rho$
is, of course, integrated out of the chiral Lagrangian which means that its
effects are incorporated into the counterterms $L_i^r$, $C^r$, etc.  Since
these counterterms do not appear in (\ref{2loop4}), the one-loop meson
contribution is suppressed.

\section*{GNC quarks}

The GNC Lagrangian\cite{GNC} contains the octet of pseudoscalar mesons
$\pi^a$ and a quark triplet $\psi$ with a dynamical quark mass
${\Sigma}(q^2)$ and a current quark mass matrix $\cal M$.
\begin{eqnarray}
{\cal L}_{GNC}(x,y) & = & \overline{\psi}(x){\delta}(x-y)
     [i\gamma^\mu(\partial_\mu-iR_\mu(y))-{\cal M}]{\psi}(y) \nonumber \\
     &  & -\overline{\psi}(x){\Sigma}(x-y){\xi}(x)X(x,y){\xi}(y){\psi}(y)
\label{sx} \\
X(x,y) & = & Pexp\left[-i\int_x^y\Gamma_{\mu}(z)dz^{\mu}\right] \\
\Gamma_\mu(z) & = & \frac{i}{2}[\xi(z)(\partial_\mu-iR_\mu(z))\xi^\dagger(z)
                    + \xi^\dagger(z)(\partial_\mu-iL_\mu(z))\xi(z)] \\
{\xi}(x) & = & exp\left[\frac{-i\gamma_5}{f_0}\sum_{a=1}^8
                          \lambda^a{\pi^a}(x)\right] \\
{\Sigma}(-q^2) & = & \frac{(A+1)m_0^3}{Am_0^2-q^2} \label{sq}
\end{eqnarray}
Note that $X(x,y)$ is a path-ordered exponential.  $L_\mu = V_\mu -
A_\mu\gamma_5$ and $R_\mu = V_\mu + A_\mu\gamma_5$ are
left and right handed external gauge fields, respectively.  For ${\cal M} =
0$
the model has local $SU(3)_L \times SU(3)_R$ symmetry, like QCD in the
presence of external gauge fields.  Numerically, we use the current quark
masses $m_u=m_d=8$ MeV and $m_s=180$ MeV.

The dynamical quark mass ${\Sigma}(-q^2)$ in (\ref{sq}) is the Fourier
transform of the ${\Sigma}(x-y)$ appearing in (\ref{sx}).  The parameter $A$
specifies the value of $m_0$ through its relation to $f_0$.
\begin{equation}
f_0^2 = \frac{N_c}{8\pi^2}\int_0^{\infty}ds\frac{s{\Sigma}(s)\left[2{\Sigma}(s)
        -s\Sigma^{\prime}(s)\right]}{\left[s+\Sigma^2(s)\right]^2}
\end{equation}
For the most part we will use the values $f_0=84$ MeV and $A=2$ which
correspond
to $m_0=317$ MeV.

The GNC quark contribution comes from two one-quark-loop contributions
to the vacuum polarization --- one with the two external photons attached at
one vertex, and the other with the two photons attached at two distinct
vertices.  The resulting contribution to $\Pi_{had}(q^2)$ is shown in
fig.~\ref{plot2}.  In this contribution we have included the small effects of
the naive one-loop graphs of the $c$ and $b$ quarks.  (The commonly-used ranges
$1.3<m_c<1.7$ GeV and $4.7<m_b<5.3$ GeV produce an uncertainty less than
2\%.)  The gluonic corrections to light and heavy quark-loop graphs will be
treated below.  We may also consider the sensitivity of the GNC result to the
quantities $f_0$ and $A$ within an allowed range  84 MeV $<f_0<$ 88 MeV and
$2<A<3$.\cite{GNC}  The GNC contribution is reduced by less than 15\% and we
find that $0.19 \lapx C^r(m_{\rho}) \lapx 0.20$~.

In fig.~\ref{plot3}, we compare our $u$, $d$, $s$ GNC result to the analogous
contribution from a more naive model with momentum independent masses of
typical values $m_u=m_d=330$ MeV and $m_s=550$ MeV.  Note that the
{\it shapes\/} of the two curves are quite different, and this is true for any
values of the momentum independent quark masses.  The GNC curve is more
consistent with the fit of Burkhardt et. al.\cite{Burkhardt}

\section*{Perturbative gluons}

The final contribution that must be considered is due to radiative gluons.
We have already included one-quark-loop diagrams with the correct non-zero
masses (momentum dependent masses for the light quarks), so we wish to
extract the gluonic corrections given by terms containing $\alpha_s$.  We
use the following dispersion relation
\begin{equation}\label{disprel}
\left[\Pi_{had}(q^2)-\Pi_{had}(0)\right]_{pert} =
       \frac{-q^2}{12\pi^2}\int_{m_c^2}^{\infty}ds\frac{R(s)-R_0(s)}{s(s-q^2)}
\end{equation}
where $q^2<0$.  $R(s)$ has been calculated using the $\overline{MS}$ scheme
to ${\cal O}((\alpha_s/{\pi})^3)$ in
\cite{Gorishny} for $N_f$ quark flavours with charges $Q_i$.
\begin{eqnarray}
   R(s) & = & 3\left({\sum}Q_i^2\right)
              \left[ 1 + \frac{\alpha_s}{\pi}
              + r_1\left(\frac{\alpha_s}{\pi}\right)^2
              + r_2\left(\frac{\alpha_s}{\pi}\right)^3 \right]
              + {\cal O}\left(\frac{\alpha_s}{\pi}\right)^4 \\
   r_1 & = & 1.9857-0.1153N_f \\
   r_2 & = & -6.6368-1.2001N_f-0.0052N_f^2
             -1.2395\left[\frac{({\sum}Q_i)^2}{3{\sum}(Q_i^2)}\right]
\end{eqnarray}
$R_0(s)$ is the
value of $R(s)$ when $\alpha_s$ is set to zero, and its appearance in
(\ref{disprel}) removes the naive one-quark-loop result.

$\alpha_s(s)$ is obtained by solving the QCD $\beta$ function\cite{Tarasov}.
\begin{eqnarray}
   \mu\frac{d\alpha_s}{d\mu} & = & -\frac{\beta_0}{2\pi}\alpha_s^2
   -\frac{\beta_1}{8\pi^2}\alpha_s^3-\frac{\beta_2}{32\pi^3}\alpha_s^4
   + {\cal O}(\alpha_s^5) \\
   \beta_0 & = & 11-\frac{2}{3}N_f \\
   \beta_1 & = & 102-\frac{38}{3}N_f \\
   \beta_2 & = & \frac{2857}{2}-\frac{5033}{18}N_f+\frac{325}{54}N_f^2
\end{eqnarray}
We have cut off the integral in (\ref{disprel}) at $s=m_c^2$.
The result is an estimate of hard gluonic corrections missing in the GNC
model, and it is shown separately in figs.~\ref{plot1} and \ref{plot2}.
The gluonic contribution contains an uncertainty of roughly 23\%(16\%) at
$-q^2=m_\tau^2(M_Z^2)$ corresponding to the range
$.11 < \alpha_s(M_Z^2) < .13$
and an uncertainty of 26\%(6\%) at  $-q^2=m_\tau^2(M_Z^2)$ for
$1.3 < m_c < 1.7$ GeV.

By our choice of the cutoff in (\ref{disprel}) we have added to the GNC quark
model only those perturbative corrections which can be reliably
calculated.  Our intent is to see how the model does at describing the
contributions that cannot be calculated perturbatively.

\section*{Lepton anomalous magnetic moments}

To lowest order in $\alpha_{QED}$ the hadronic
contribution to a lepton's magnetic moment has the following algebraic
form\cite{Lifshitz}.
\begin{eqnarray}\label{magmom1}
a^{(4)}_{l,had} & = & \left(\frac{g-2}{2}\right)^{(4)}_{l,had}
             = \frac{1}{\pi}\int_{4m_l^2}^{\infty}\frac{dt}{t}X_l(t) \\
X_l(t) & = & \frac{e^4m_l^2}{4\pi\sqrt{t(t-4m_l^2)}}\int_{-1}^1dz\left(
             \frac{1+3z}{2}\right){\Pi}_{had}(f_l^2) \\
f_l^2 & = & -\frac{1}{2}(t-4m_l^2)(1-z)
\label{magmom3}
\end{eqnarray}
The superscript ``(4)'' reminds us that this is all at ${\cal O}(e^4)$.  Notice
that $f_l^2
\leq 0$ for the entire range of integration, so we only need
the value of $\Pi_{had}(q^2)$ for spacelike $q^2$.

We perform the double integration by fitting our numerical result for
$\Pi_{had}(q^2)$ to a piecewise-analytic function of $q^2<0$.  Since $m_\mu
\ll m_c$, our estimated $\alpha_s$ corrections to $\Pi_{had}(q^2)$ have
essentially no effect on $a^{(4)}_{\mu,had}$. For the case of
$a^{(4)}_{\tau,had}$ the uncertainties in the meson-loop contribution and
the $\alpha_s$ corrections are both larger, with the former dominating.  We
take a 40\% error for the meson-loop contribution in this case, double the
naive error due to the uncertainty in the $L_i^r$'s as noted above, to
include possible corrections from higher order effects in the chiral
Lagrangian.  The results are \begin{eqnarray}\label{c1}
a^{(4)}_{\mu,had} & = & (6.3 \pm 0.5) \times 10^{-8} \\
a^{(4)}_{\tau,had} & = & (3.2 \pm 0.1) \times 10^{-6}
\end{eqnarray}
Any other choice of ($f_0$,$A$) in the range (84 MeV $<f_0<$ 88 MeV, $2<A<3$)
would reduce our result for $a^{(4)}_{\mu,had}$($a^{(4)}_{\tau,had}$) by
less than 10\%(6\%).

These results should be compared to the calculation from integrating an
experimental determination of $R(q^2)$.  ($R(q^2)$ is defined in
(\ref{Rratio}).  The precise form of the integral is given by Kinoshita et.
al.\cite{Martinovic})
\begin{equation}\label{expt}
{\rm experiment} \Rightarrow
  \left\{
  \begin{array}{llll}
  a^{(4)}_{\mu,had} & = & (7.05 \pm 0.08) \times 10^{-8} & \cite{Martinovic} \\
  a^{(4)}_{\tau,had} & = & (3.6 \pm 0.3) \times 10^{-6} & \cite{Samuel}
  \end{array}
  \right.
\end{equation}

On the other hand the parametrization
of the experimentally-determined $\Pi_{had}(q^2)$ due to
Burkhardt et. al.\cite{Burkhardt} may be used directly in
eqs.~(\ref{magmom1}-\ref{magmom3}).  This gives
\begin{equation}
{\rm experimental~fit} \Rightarrow
   \left\{
   \begin{array}{lll}
   a^{(4)}_{\mu,had} & = & 6.63 \times 10^{-8} \\
   a^{(4)}_{\tau,had} & = & 3.45 \times 10^{-6}
   \end{array}
   \right.
\end{equation}
The authors of \cite{Burkhardt} seem to claim that the uncertainty should
be less than 5\%, making this result for $a^{(4)}_{\mu,had}$ noticeably smaller
than the preceding result of (\ref{expt}).

Finally, we consider quarks with momentum independent masses of
$m_u=m_d=330$ MeV and
$m_s=550$ MeV, and add the pseudoscalar mesons, heavy quarks and $\alpha_s$
corrections exactly as discussed above.
\begin{equation}
m_u=m_d=330 {\rm MeV},m_s=550 {\rm MeV} \Rightarrow
   \left\{
   \begin{array}{lll}
   a^{(4)}_{\mu,had} & = & (4.3 \pm 0.5) \times 10^{-8} \\
   a^{(4)}_{\tau,had} & = & (2.2 \pm 0.1) \times 10^{-6}
   \end{array}
   \right.
\end{equation}
Clearly the GNC model is a significant improvement.  In order to reproduce
the GNC results for both lepton magnetic moments in
(\ref{c1}) we would require {\it different\/} sets of masses, for example
$m_{u,d,s}= 243$ MeV and $m_{u,d,s}= 201$ MeV
for $a^{(4)}_{\mu,had}$ and $a^{(4)}_{\tau,had}$
respectively.  This illustrates the fact mentioned previously that momentum
independent quark masses cannot reproduce the shape of our $\Pi_{had}(q^2)$
for the entire range $-q^2<0$.

\section*{QED running coupling}

It is straightforward to evaluate the hadronic contribution to the running of
$\alpha_{QED}(-q^2)$ from our results.\cite{alpha}  Using $f_0=84$ MeV and
$A=2$ we find
\begin{equation} \label{ours}
\Delta \equiv
\left[\frac{1}{\alpha_{QED}(0)}-\frac{1}{\alpha_{QED}(M_Z^2)}\right]_{had} =
4{\pi}[\Pi_{had}(-M_Z^2)-\Pi_{had}(0)] = 3.68 \pm 0.07
\end{equation}
The error is dominated by the error in the perturbative QCD contribution.
This may be compared to the result in \cite{Burkhardt}.
\begin{equation}
\Delta = 3.94 \pm 0.12
\end{equation}
When combined with the well-known non-hadronic effects, our result (\ref{ours})
implies
\begin{equation}
\alpha_{QED}^{-1}(M_Z^2) = 129.05 \pm 0.07
\end{equation}
We note that this is also consistent with an independent determination using
recent LEP data.\cite{Banerjee}  (In that analysis additional corrections are
included to define an effective coupling, $\alpha_{eff}(M_Z)$).
Any other choice of ($f_0$,$A$) in the range (84 MeV $<f_0<$ 88 MeV,
$2<A<3$) would reduce the value of $\Delta$ by less than 2\%.

\section*{Comments}

In this paper we have used a nonlocal constituent quark model for the
description of $\Pi_{had}(q^2)$ at spacelike momenta.  The success of our
quark level description could be considered to be a manifestation of
``duality'', and it clearly relies on being far removed from the resonance
structure of QCD appearing for timelike momenta.  The various uncertainties
we have quoted occur within the model itself, and they are not intended as
an {\em a priori} estimate of how well the model should resemble QCD.  It is
only after a comparison with experimental data that our model is able to
shed light on some of the essential features of QCD dynamics.

After completion of this work, we received a preprint\cite{deRafael}
containing an independent theoretical estimate of the hadronic vacuum
polarization and the muon magnetic moment.

\section*{Appendix}

The complete result for $\Pi_{had}(q^2)$ for small $-q^2$ can be
derived from the chiral Lagrangian, and is given in (\ref{2loop}).  Here we
provide the explicit form of the $T_i$ parameters, using the notation of
\cite{G&L} at the renormalization scale $\mu$.  (The $\mu$ and $q^2$
dependence of all parameters is implicit in this appendix to simplify the
notation.)  The parameters ${\tilde m}_\pi$, ${\tilde m}_K$ and
${\tilde m}_\eta$ are the {\it lowest order} mass values.
The relation of these parameters to the physical masses can be taken from
eq.~(10.7) of \cite{G&L}.  The $T_i$'s are well-behaved functions of
spacelike $q^2$ such that $q^2T_i$ vanishes when $q^2=0$.  The apparent
$\mu$ dependence in $T_2$, $T_3$, and  $T_4$ cancels out.
\begin{eqnarray}
T_1({\mu}) & = & 8l_\pi+8l_K
       -\frac{2}{3(4{\pi})^2}\left( l_\pi^2+l_K^2+l_{\pi}l_K \right) \\
T_2 & = & \frac{8}{3}\left[L_9^r
          -\frac{1}{8(4{\pi})^2}\left( \frac{1}{6}+l_\pi+l_K \right)\right] \\
T_3(q^2) & = & Y_1\phi_\pi + Y_2\phi_K \nonumber \\
    &   & + 8\frac{{\tilde m}_\pi^2}{q^2}\left[2L_4^r+L_5^r-4L_6^r-2L_8^r
          -\frac{1}{24(4{\pi})^2}\left(3l_\pi-\frac{1}{3}l_\eta\right)\right]
          \nonumber \\
    &   & + 8\frac{{\tilde m}_K^2}{q^2}\left[4L_4^r+L_5^r-8L_6^r-2L_8^r
          -\frac{1}{18(4{\pi})^2}l_\eta\right] \\
T_4(q^2) & = & Y_3\phi_\pi^2 + Y_4\phi_\pi\phi_K + Y_5\phi_K^2 \\
Y_1 & = & 4\left(1-4\frac{{\tilde m}_\pi^2}{q^2}\right)\left[L_9^r-
          \frac{1}{12(4\pi)^2}\left(\frac{1}{2}+2l_\pi+l_K\right)\right]
          \nonumber \\
    &   & +48\frac{{\tilde m}_\pi^4}{q^4}\left[L_4^r+L_5^r-2L_6^r-2L_8^r-
          \frac{1}{24(4\pi)^2}\left(3l_\pi+\frac{1}{3}l_\eta\right)\right]
          \nonumber \\
    &   & +96\frac{{\tilde m}_\pi^2{\tilde m}_K^2}{q^4}\left[L_4^r-2L_6^r+
          \frac{1}{36(4\pi)^2}l_\eta\right] \\
Y_2 & = & 4\left(1-4\frac{{\tilde m}_K^2}{q^2}\right)\left[L_9^r-
          \frac{1}{12(4\pi)^2}\left(\frac{1}{2}+l_\pi+2l_K\right)\right]
          \nonumber \\
    &   & +48\frac{{\tilde m}_K^4}{q^4}\left[2L_4^r+L_5^r-4L_6^r-2L_8^r-
          \frac{1}{9(4\pi)^2}l_\eta\right]
          \nonumber \\
    &   & +48\frac{{\tilde m}_\pi^2{\tilde m}_K^2}{q^4}\left[L_4^r-2L_6^r+
          \frac{1}{36(4\pi)^2}l_\eta\right] \\
Y_3 & = & \frac{-1}{6(4{\pi})^2}\left(1-4\frac{{\tilde m}_\pi^2}{q^2}\right)^2
          \\
Y_4 & = & \frac{-1}{6(4{\pi})^2}\left(1-4\frac{{\tilde m}_\pi^2}{q^2}\right)
           \left(1-4\frac{{\tilde m}_K^2}{q^2}\right) \\
Y_5 & = & \frac{-1}{6(4{\pi})^2}\left(1-4\frac{{\tilde m}_K^2}{q^2}\right)^2 \\
l_P & = & ln(\frac{{\tilde m}_P}{\mu}) \\
\phi_P & = & \int_0^1dz~ln\left(1-z(1-z)\frac{q^2}{{\tilde m}_P^2}\right) \\
P & = & (\pi,K,{\eta})
\end{eqnarray}

\section*{Acknowledgements}

R.R.M. benefited from conversations with M. Ahmady, V. Elias, N. Hill,
M. A. Samuel and T. G. Steele.  R.L. would like to thank T. Morozumi,
M. Sutherland and C. Wicentowich for helpful discussions.
This research was supported in part by the Natural Sciences and
Engineering Research Council of Canada.



\section*{Figure Captions}

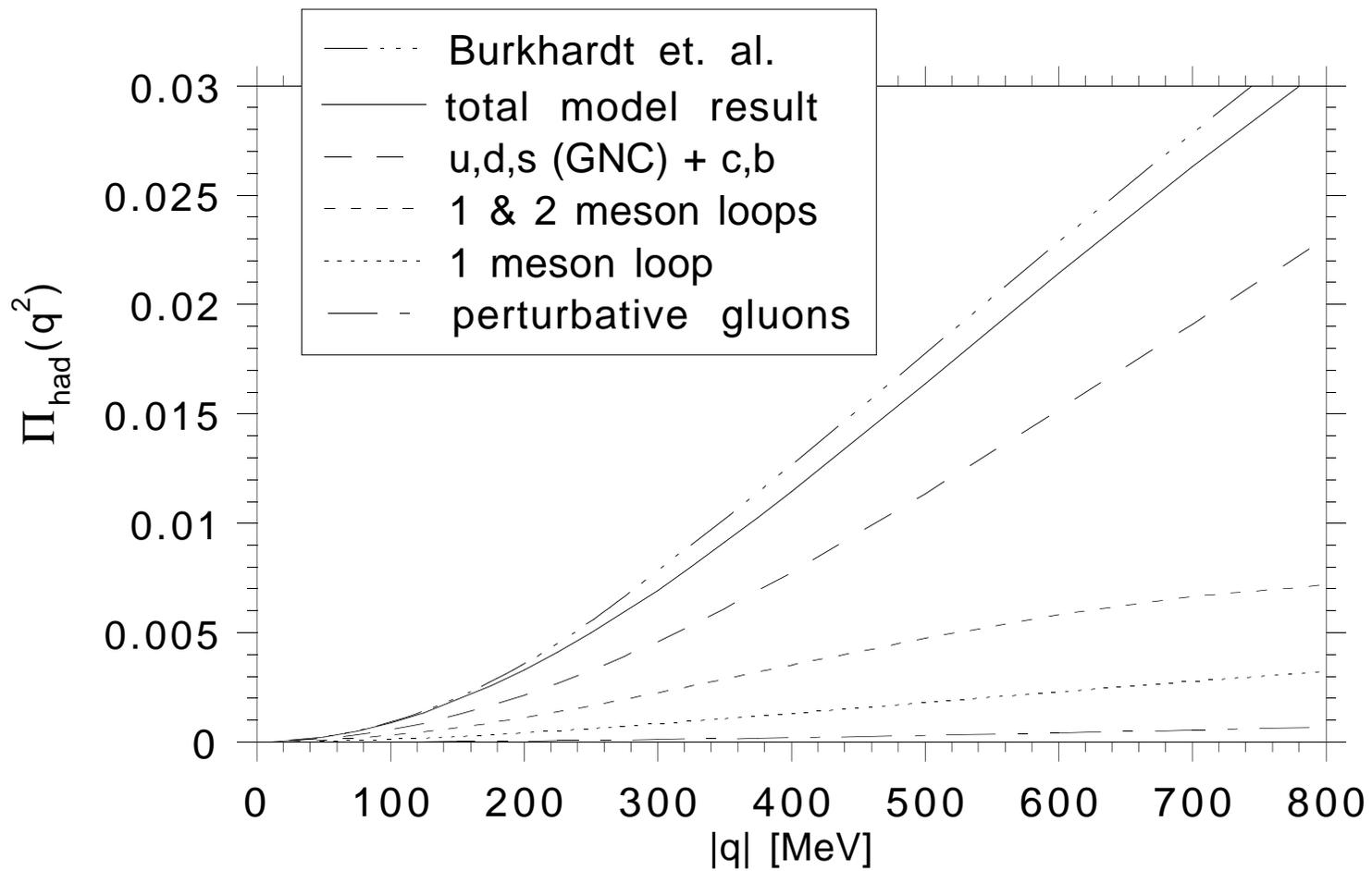
\begin{figure}[h]
\caption{All contributions to our total result for $\Pi_{had}(q^2)$ for
         $|q|<800$MeV.  The fit from
         \protect\cite{Burkhardt} is also shown.}\label{plot1}
\end{figure}
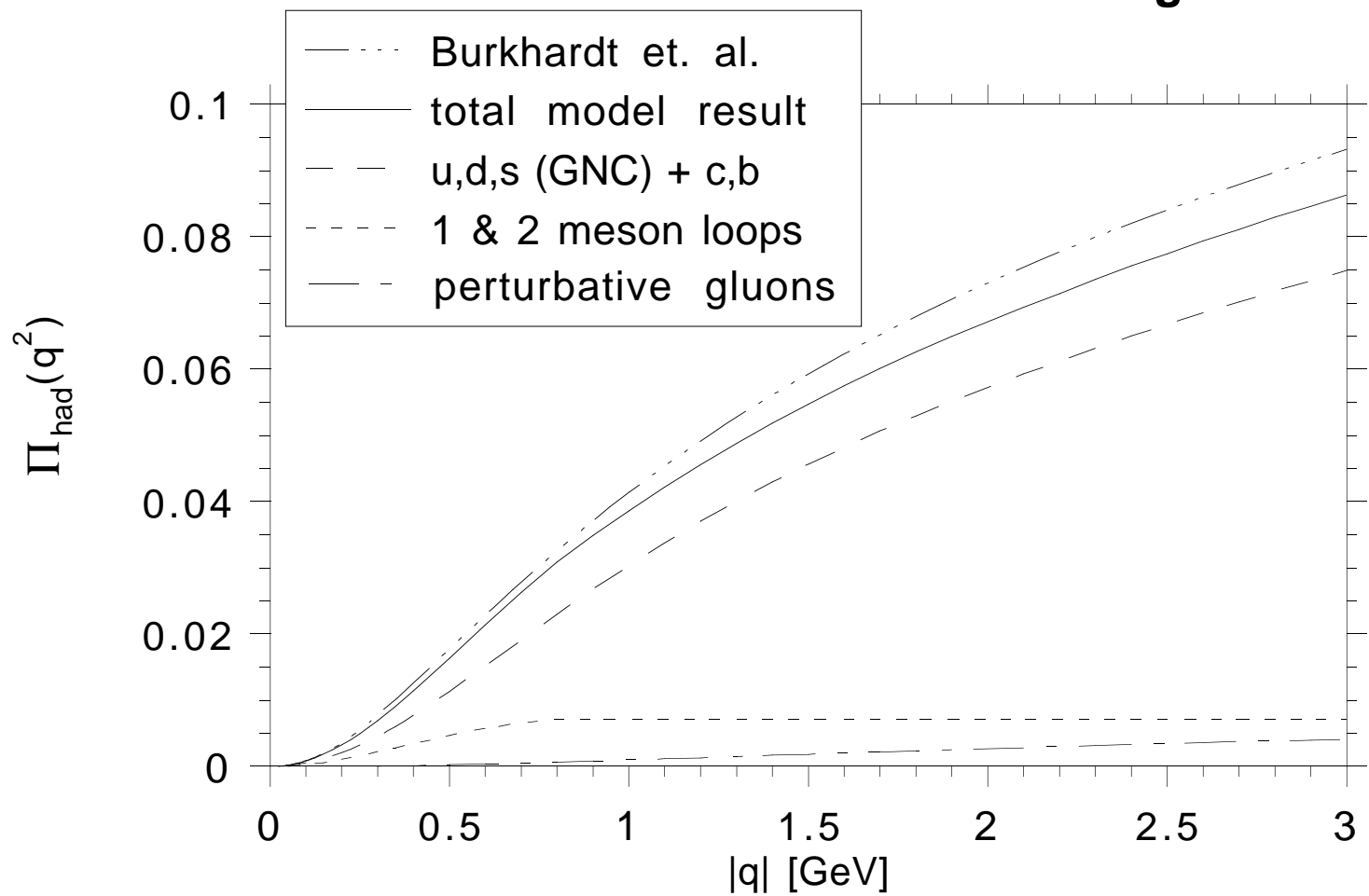
\begin{figure}[h]
\caption{All contributions to our total result for $\Pi_{had}(q^2)$ for
         $|q|<3$GeV.
         The fit from \protect\cite{Burkhardt} is also shown.}\label{plot2}
\end{figure}
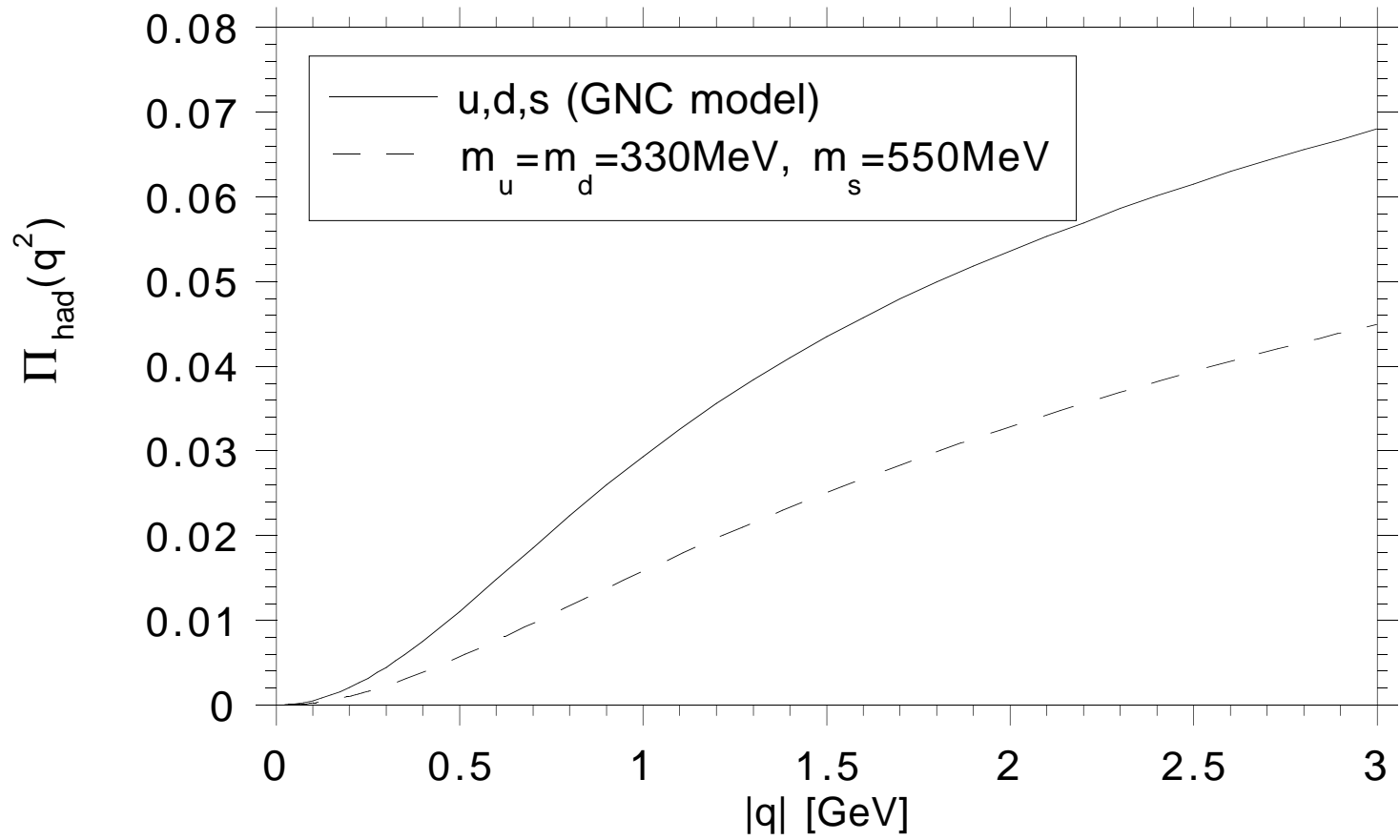
\begin{figure}[h]
\caption{Comparison of the contribution to $\Pi_{had}(q^2)$ from the $u$, $d$
         and $s$ quarks in
         two distinct approaches:  the GNC model (as used in the present
         analysis) and a simple model with momentum independent masses
         $m_u=m_d=330$ MeV and $m_s=550$ MeV.}\label{plot3}
\end{figure}

\end{document}